\documentclass[lettersize,journal]{IEEEtran}
\usepackage{amsmath,amsfonts}
\usepackage{amssymb}
\usepackage{algorithmic}
\usepackage{algorithm}
\usepackage{array}
\usepackage{textcomp}
\usepackage[caption=true,font=normalsize,labelfont=sf,textfont=sf]{subfig}
\usepackage[colorlinks,linkcolor=blue,]{hyperref}
\usepackage{stfloats}
\usepackage{graphicx}
\usepackage{wrapfig}
\usepackage{url}
\usepackage{verbatim}
\usepackage{graphicx}
\usepackage{cite}
\usepackage{bm}
\usepackage{xcolor}
\usepackage{multirow}
\usepackage{booktabs}
\usepackage{amsfonts}
\usepackage{enumerate}
\usepackage{siunitx}
\usepackage{gensymb}
\usepackage{bbding}
\usepackage{soul} 
\usepackage{color, xcolor}
\usepackage{afterpage}

\begin{document}

\title{Multimodal Fish Feeding Intensity Assessment in Aquaculture}

\author{Meng Cui, Xubo Liu, Haohe Liu,  Zhuangzhuang Du, Tao Chen,
Guoping Lian, Daoliang Li, Wenwu Wang

\thanks{M. Cui, X. Liu, H. Liu, and W. Wang are with the Centre for Vision, Speech and Signal Processing (CVSSP), University of Surrey, Guildford GU2 7XH, UK. (e-mail:
[m.cui, xubo.liu, haohe.liu, w.wang]@surrey.ac.uk).}
\thanks{T. Chen and G. Lian are with the Department of Chemical and Process Engineering, University of Surrey, Guilford GU2 7XH, UK. (e-mail: [t.chen, g.lian]@surrey.ac.uk).}
\thanks{Z. Du and D. Li are with the National Innovation Center for Digital Fishery, China Agricultural University, China (e-mail: 18437956159@163.com, dliangl@cau.edu.cn).}
}

\maketitle
\begin{abstract}

Fish feeding intensity assessment (FFIA) aims to evaluate fish appetite changes during feeding, which is crucial in industrial aquaculture applications. Existing FFIA methods are limited by their robustness to noise, computational complexity, and the lack of public datasets for developing the models. To address these issues, we first introduce \textit{AV-FFIA}, a new dataset containing 27,000 labeled audio and video clips that capture different levels of fish feeding intensity. Then, we introduce multi-modal approaches for FFIA by leveraging the models pre-trained on individual modalities and fused with data fusion methods. We perform benchmark studies of these methods on AV-FFIA, and demonstrate the advantages of the multi-modal approach over the single-modality based approach, especially in noisy environments. However, compared to the methods developed for individual modalities, the multimodal approaches may involve higher computational costs due to the need for independent encoders for each modality. To overcome this issue, we further present a novel unified mixed-modality based method for FFIA, termed as \textit{U-FFIA}. U-FFIA is a single model capable of processing audio, visual, or audio-visual modalities, by leveraging modality dropout during training and knowledge distillation using the models pre-trained with data from single modality. We demonstrate that U-FFIA can achieve performance better than or on par with the state-of-the-art modality-specific FFIA models, with significantly lower computational overhead, enabling robust and efficient FFIA for improved aquaculture management. To encourage further research, we have released the AV-FFIA dataset, the pre-trained model and codes at \url{https://github.com/FishMaster93/U-FFIA}.

\end{abstract}

\def\abstractname{Note to Practitioners}
\begin{abstract}
Feeding is one of the most important costs in aquaculture. However, current feeding machines usually operate with fixed thresholds or human experiences, lacking the ability to automatically adjust to fish feeding intensity. FFIA can evaluate the intensity changes in fish appetite during the feeding process and optimize the control strategies of the feeding machine to avoid inadequate feeding or overfeeding, thereby reducing the feeding cost and improving the well-being of fish in industrial aquaculture. The existing methods have mainly exploited single-modality data, and have a high sensitivity to input noise. Using video and audio offers improved chances to address the challenges brought by various environments. However, compared with processing data from single modalities, using multiple modalities simultaneously often involves increased computational resources, including memory, processing power, and storage. This can impact system performance and scalability. To address these issues, we focus on the efficient unified model, which is capable of processing both multimodal and single-modal input. Our proposed model achieved state-of-the-art (SOTA) performance in FFIA with high computational efficiency.
\end{abstract}
\begin{IEEEkeywords}
Fish feeding intensity assessment (FFIA), computer vision, acoustic technology, audio-visual fusion

\end{IEEEkeywords}
\section{Introduction}

\IEEEPARstart
{D}{igital} aquaculture plays an important role in enhancing operational efficiency, sustainability, and productivity in aquaculture \cite{wang2021intelligent}. As the foundation of digital aquaculture technology, \textcolor{black}{fish feeding intensity assessment (FFIA) is the task of evaluating the fish feeding intensity during the feeding process \cite{yang2021deep}.} FFIA offers significant benefits for real-world aquaculture by facilitating adjustments to feed dispensing, minimizing waste, and ultimately improving productivity \cite{atoum2014automatic}.

Traditional FFIA methods have relied on human observers' experience and subjective interpretation \cite{ an2021survey}, which can be time-consuming and labour-intensive. \textcolor{black}{With the recent advances in deep learning, data-driven approaches  \cite{zhou2019evaluation} have emerged as promising methods for FFIA, such as image-based fish feeding intensity classification} (e.g., ``\textit{None}'', ``\textit{Weak}'', ``\textit{Medium}'' and ``\textit{Strong}'' \cite{ubina2021evaluating}) using convolutional neural networks (CNNs) \cite{cheng2022convolutional}, and video-based FFIA methods \cite{wang2021automatic, maaloy2019spatio}. Video-based methods offer advantages in capturing spatial and temporal information of the feeding process, providing a richer context for fish feeding behaviour analysis compared to single-image methods \cite{feng2022fish}. However, processing video data often requires greater computational resources, which may make it less practical for on-device applications that play a crucial role in aquaculture \cite{li2020automatic}. Furthermore, vision-based methods are generally susceptible to variations in lighting conditions and noise from water surface reflections \cite{zhou2017near}, affecting their reliability in real-world aquaculture settings.

As an alternative to vision-based methods, acoustic measurements can also be used for FFIA. In comparison, acoustic measurements are unaffected by illumination changes and occlusions, suitable for round-the-clock 24 hours monitoring, and are often more cost-effective to process \cite{gao2020listen, lin2022eclipse}. However, acoustics-based methods can be limited in capturing the physical characteristics related to fish behaviour, and are prone to underwater acoustic noises, potentially compromising FFIA performance.

Exploiting the complementarity between audio and vision has the potential to address their respective limitations. However, its potential has not been explored in FFIA. In this work, we introduce audio-visual (AV) approaches for FFIA by leveraging pre-trained models for each modality and the methods for fusing the modalities. We show the advantages of the multi-modal approaches over the single-modal approaches, especially in noisy environments. We also demonstrate that with the multimodal approaches, the computational cost may be increased due to the need for processing the data from the individual modalities simultaneously. In addition, the absence of an individual modality can degrade the performance of the overall model.

To further improve the robustness and efficiency of multimodal FFIA, we propose a novel unified mixed-modality model, termed as \textit{U-FFIA}, which is the first model that is capable of processing audio, visual or audio-visual modalities in the FFIA context. Our method encapsulates three main innovations. First, modality dropout is used to fuse information with missing modalities. Second, we use SimPFs \cite{liu2023simple} to reduce the redundant information within the mel-spectrogram and use the pre-trained audio models \cite{kong2020panns} to extract audio features. Third, we use the knowledge distillation method \cite{hinton2015distilling}, taking a pre-trained model with full audio and video frames as the teacher network, to enhance the performance of the audio and video student network with a reduced number of frames. 
 
To address the dataset scarcity issue, we also developed a large-scale audio-visual dataset for FFIA, called \textit{AV-FFIA}, which comprises 27, 000 labelled audio and video clips. To our knowledge, AV-FFIA is the first large-scale multimodal dataset for FFIA research. Compared to the existing \textit{AFFIA3K} dataset \cite{cui2022fish}, AV-FFIA is approximately \num{9} times larger and includes video clips corresponding to the audio clips.

This paper extends our previous work presented at a conference \cite{cui2022fish}, with the following new contributions. (1) We propose the first audio-visual approach for FFIA, which significantly enhances the robustness of the recognition system operating in noisy environments (e.g., audio bubbles noise, visual corruption).
(2) We propose U-FFIA, a novel unified model that achieves state-of-the-art performance on the FFIA task while efficiently processing both multimodal and single-modal inputs.
(3) We introduce AV-FFIA, the first large-scale multimodal dataset for FFIA and conduct comprehensive studies of the performance of the proposed U-FFIA, as compared with the audio-only, vision-only, and audio-visual benchmarking models. 

This paper is organized as follows: The related works about FFIA are described in Section \ref{sec: related work}. Section \ref{sec:benchmark} describes our benchmark studies on several audio-visual fusion-based deep learning frameworks for FFIA. Our proposed novel U-FFIA methods are described in Section \ref{sec: U-FFIA}. Section \ref{sec:dataset} introduces audio and video datasets of FFIA that we have captured. 
Experimental details are shown in Section \ref{sec:experiment}. Section \ref{sec:results} presents the experimental results and provides an in-depth analysis of the factors contributing to the results. Section \ref{sec:conclusion} concludes the contribution of this paper and discusses the future direction.

\section{Related work}
\label{sec: related work}

Vision-based FFIA methods have initially considered single images, and employed traditional techniques, such as background segmentation and target tracking \cite{zhao2017assessing}. However, these methods suffered from high computational loads and were prone to environmental factors like water surface fluctuations \cite{zhou2017near}.
More recent works have considered recognition tasks including fish classification \cite{mathur2020crosspooled}, counting \cite{albuquerque2019automatic}, tracking \cite{liu2018embedded, wang2021daniosense} and trawl fishing \cite{gu2024mfgtn} by leveraging CNN or Transformer models. For FFIA, models such as MSIF-MobileNetV3 \cite{zhang2023msif} and EfficientNet-B2 \cite{yang2021dual} have been successfully applied. 

However, these image-based methods have limitations in capturing the dynamic and continuous nature of fish feeding behaviour \cite{maaloy2019spatio}. To address this, video-based methods that capture both spatial and temporal information have been proposed \cite{ su2020visual}. A common approach is to convert the original RGB video into an optical flow image sequence and feed it into a 3D CNN, which has been shown to outperform image-based models \cite{ wang2021automatic, ubina2021evaluating}. Despite their effectiveness, video-based methods are computationally demanding and face challenges such as uneven illumination and occlusions.

As an alternative to vision-based methods, audio-based FFIA methods leverage the vocalizations produced by fish during the feeding process, where the feeding sounds vary with the intensity of food intake, allowing for the analysis of feeding behaviour through the frequency spectrum \cite{phillips1989feeding, yamaguchi1975spectrum, shishkova1958notes, takemura1988attraction, qi2023effects}.
Audio-based FFIA was initially proposed by \cite{cui2022fish}, where the audio signal is first transformed into log mel spectrograms and then fed into a CNN-based model for FFIA. Subsequent work \cite{du2023feeding, zeng2023fish} have further demonstrated the feasibility of using audio for FFIA. Audio-based methods offer advantages such as energy efficiency and lower computational costs compared to vision-based methods \cite{gao2018learning, gao2020listen}. However, audio-based models have lower classification performance compared to video-based FFIA due to their inability to capture full visual information and sensitivity to environmental noise \cite{choi2022temporal}

The fusion of audio and visual modalities has the potential to address the limitations posed by each modality. This has been demonstrated in various fields, such as audio-visual event localization \cite{wu2019dual}, audio-visual speech recognition \cite{li2022recent}, audio-visual emotion recognition \cite{hossain2019emotion}, and animal behaviour analysis \cite{ bold2019bird, li2023long}.
However, to the best of our knowledge, no previous work has explored the fusion of audio-visual information for FFIA, which is the focus of our work in this paper.

Processing multiple modalities simultaneously, however,
requires increased computational resources. To address those issues, recent work has proposed using a single model to process different modalities, such as SkillNet \cite{dai2022one}, EAO \cite{shvetsova2022everything}, data2vec \cite{baevski2022data2vec}, UVAM \cite{gong2022uavm}, and U-Hubert \cite{hsu2022u}. These unified models have shown promising results in efficiently combining information from multiple modalities. 
However, these models are often too large and computationally expensive to be applied on devices with limited resources, which is a common requirement in real-world aquaculture settings. In addition, no existing work has explored the use of a unified model in the context of FFIA. To bridge this gap, we propose a novel unified model U-FFIA, specially designed for the FFIA task, as discussed in subsequent sections.

\section{Audio-Visual FFIA}
\label{sec:benchmark}

To exploit the complementarity between audio and visual modalities, this section presents methods for audio-visual FFIA, with a focus on various strategies for fusing the audio and visual modalities. This will also allow us to conduct extensive benchmark studies on our AV-FFIA dataset, to be presented in Section \ref{sec:dataset}, and to investigate the potential for improving FFIA accuracy and robustness in aquaculture by fusing the audio and visual modalities, as compared with audio-only and video-only based methods.

\subsection{AV Fusion via Multi-Head Self-Attention}
The multi-head self-attention fusion method is a powerful approach that leverages the conventional Transformer architecture \cite{vaswani2017attention} for processing multimodal inputs.  In this method, each video frame and mel-spectrogram are independently split into patches and encoded using the approach proposed in ViT \cite{dosovitskiy2020image}. The encoded audio and video tokens are then concatenated to form a new sequence of tokens, which serves as the input to the multi-head attention (MHA) module. The MHA operation can be formulated as follows:

\begin{equation}
\begin{aligned}
\operatorname{MHA}(\mathbf{Q}, \mathbf{K}, \mathbf{V})=\operatorname{Softmax}\left(\frac{\mathbf{Q K}^{\top}}{\sqrt{d}}\right) \mathbf{V}
\end{aligned}
\end{equation}
where $\mathbf{Q}, \mathbf{K}, \mathbf{V}$ are the query, key, and value matrices obtained using learnable projection weights $\mathbf{W}^Q, \mathbf{W}^K, \mathbf{W}^V \in \mathbb{R}^{d \times d}$ respectively.

\subsection{AV Fusion via Multi-Head Cross-Attention}
In contrast to self-attention, multi-head cross-attention (MHCA) \cite{zhao2024cross, zheng2024multi} allows for the integration of both audio and visual information, leveraging the strengths of each modality. The MHCA operation can be written as:

\begin{equation}
\mathbf{O}^{(\ell)}=\operatorname{MHA}\left(\mathbf{V}^{(\ell-1)}, \mathbf{A}^{(\ell-1)}, \mathbf{A}^{(\ell-1)}\right)
\end{equation}
where $\mathbf{V}^{(\ell-1)}$ and $\mathbf{A}^{(\ell-1)}$ are the vision and audio patch representation at layer ${\ell-1}$. In MHCA, the learnable projection weights derived from the video patches are used as queries, while the learnable projection weights from the audio patches are used as keys and values. The MHCA module then calculates the new audio-visual fused representation $\mathbf{O}^{(\ell)}$ as a weighted summation of the audio-visual features.

\subsection{AV Fusion via Attention Bottlenecks}
Bottleneck attention \cite{nagrani2021attention} is a technique used in multimodal fusion to address the problem of attention bottlenecks, which can occur when the number of modalities or the dimensionality of the feature space is large. It can be represented as follows:
 \begin{equation}
\mathrm{z}_{\mathrm{f}}=\left[\mathrm{z}_{\mathrm{f}}^1, \mathrm{z}_{\mathrm{f}}^2, \ldots, \mathrm{z}_{\mathrm{f}}^B\right]
\end{equation}
where $\mathrm{z}_{\mathrm{f}}$ represents a set of fused bottleneck tokens, and $B$ represents the number of fused bottleneck tokens. In our experiments, we have $B=2$, and the dimension of each bottleneck token is set as \num{768}.

We then restrict all cross-modal attention flow via these bottleneck tokens. More formally, for layer $l$, we compute the token representations as follows:
\begin{equation}
\begin{aligned}
& {\left[\mathbf{z}_{\mathrm{v}}^{l+1} \| \hat{\mathbf{z}}_{\mathrm{f}}^{l+1}\right]=\operatorname{Transformer}\left(\left[\mathbf{z}_{\mathrm{v}}^l \| \mathbf{z}_{\mathrm{f}}^l\right] ; \theta_{\mathrm{v}}\right)} \\
& {\left[\mathbf{z}_{\mathrm{a}}^{l+1} \| \mathbf{z}_{\mathrm{f}}^{l+1}\right]=\operatorname{Transformer}\left(\left[\mathbf{z}_{\mathrm{a}}^l \| \hat{\mathbf{z}}_{\mathrm{f}}^{l+1}\right] ; \theta_{\mathrm{a}}\right)}
\end{aligned}
\end{equation}
where $\operatorname{Transformer}$ is the $\operatorname{Transformer}$ encoder\cite{vaswani2017attention}, 
$\mathbf{z}_{\mathrm{v}}^{l}$ is video tokens and the $\mathbf{z}_{\mathrm{a}}^{l}$ is audio spectrogram tokens. $\left[\mathbf{z}_{\mathrm{v}}^l \| \mathbf{z}_{\mathrm{f}}^l\right]$ is the concatenation of the video and bottleneck tokens and
$\left[\mathbf{z}_{\mathrm{a}}^l \| \hat{\mathbf{z}}_{\mathrm{f}}^{l+1}\right]$ is the concatenation of the audio tokens and bottleneck tokens (after being fused with the video tokens).
$\theta_{\mathrm{v}}$ represents the parameters of the Transformer encoder with the fused bottleneck and video tokens as inputs. $\theta_{\mathrm{a}}$ represent the parameters of Transformer encoders with fused bottleneck and audio tokens as inputs. Then the attention mechanisms are used to integrate the information from different modalities.

In this case, the audio and video tokens can only exchange information via the fused bottleneck token $\hat{\mathbf{z}}_{\mathrm{f}}^{l+1}$ within a $\operatorname{Transformer}$ layer.

\section{Unified Mixed-Modality Based FFIA (U-FFIA)}
\label{sec: U-FFIA}

\begin{figure*}[t]
    \centering
    \includegraphics[scale=0.5]{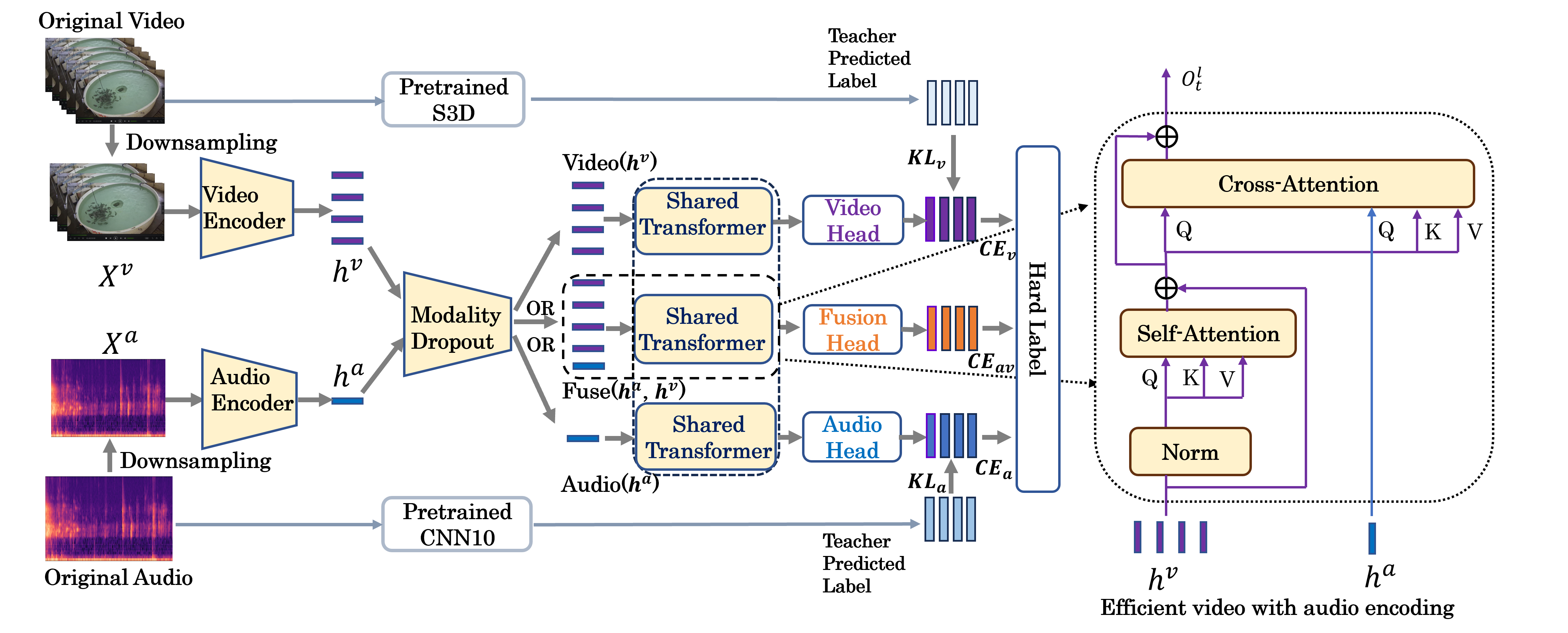}
    \caption{Overview of our proposed method. We first downsample the original video and audio, and then use a video encoder and audio encoder to extract the video and audio features, respectively. The audio encoder is a pre-trained MobileNetV2 and the video encoder is simply a linear projection layer.
 We cut the video features into non-overlap patches \num{16} $\times$ \num{16}, and use whole audio features. We use a modality dropout to randomly select one modality on each step during the model training. The shared Transformer encoder has \num{6} layers with \num{8} heads, embedding dimension \num{768}, and FFN dimension \num{1024}, using the pre-norm residual connection setup. For single-modality tasks, we employ knowledge distillation with pre-trained S3D (video) and CNN10 (audio) as teachers. For audio-visual fusion, we use cross-attention to incorporate audio cues into video frame representations.}
    \label{fig:overview}
\end{figure*}

To further improve AV-FFIA performance, we propose the U-FFIA model, as shown in Fig. \ref{fig:overview}, which includes several key components, such as preprocessing of the audio-visual inputs, audio-visual fusion, modality dropout, and knowledge distillation. These components work together to improve the model's robustness, computational efficiency, and performance. We discuss the details of these components in the following subsections.

\subsection{Audio-Visual Input}
We first use the audio pre-processing method to obtain the Mel spectrogram features of the audio signal. However, in the aquaculture environment, various sources of noise, such as bubbles and machine noise, can introduce redundancy in the mel-spectrogram data when used as input to CNNs. To tackle this issue and optimize computational efficiency, we adopt the simple pooling front-ends (SimPFs) \cite{liu2023simple, ccayli2023knowledge} method to reduce the redundant information present in the input mel-spectrograms, thereby enhancing the overall computational efficiency of the model. The spectral pooling method of SimPF computes the discrete Fourier transform (DFT) of the audio frames $\mathbf{X}$ and then crops the centre to get $\mathbf{F}^{crop}$ using a bounding box of the shape $\left(S, k N^a\right)$, where $S$ refers to the dimension of the spectral feature, $k$ denotes the compression rate, ranging from 0 to 1, and $N^a$ is the number of audio frames. Then the output of the inverse discrete Fourier transform (IDFT) ${\mathbf{C}}$ is taken as the compressed audio, as follows,

\begin{equation}
\begin{aligned}
\label{SimPF}
& \mathbf{F}=\operatorname{DFT}\left(\mathbf{X}\right) \\
& \mathbf{F}^{crop}=\mathbf{F}\left(S, k N^a\right) \\
& \mathbf{C}=\operatorname{IDFT}\left(\mathbf{F}^{crop}\right)
\end{aligned}
\end{equation}

We use the MobileNetV2 model, pre-trained on AudioSet, as the audio encoder to extract the audio features from the preprocessed mel spectrogram by the SimPFs. Afterwards, an average pooling is applied along the frequency dimension and then maximum and average operations are used along the time dimension. We sum the maximized and averaged features and then project them into the shared embedding space through a new multi-layer perception (MLP) block with two linear layers and a ReLU activation hidden layer. Finally, we obtain the audio input $ \mathbf{h}^a \in \mathbb{R}^{T_a \times d}$, where ${T_a}$ is the number of audio feature vectors, and ${d}$ is the dimension of each feature vector. A specialized audio class token $\mathbf{A}_{c l}^{(0)} \in \mathbb{R}^{ 1 \times 768}$ is added to the audio-embedded patches.

For the video modality, we randomly downsample the video frames from the entire input video (50 frames) to obtain the video input of $X^v \in \mathbb{R}^{N_f \times 3 \times H \times W}$, where $N_f$ is the number of frames, $H$ is the height of the frame, and $W$ is the width of the frame. Following the ViT approach \cite{dosovitskiy2020image}, we split each frame into $N$ no-overlapping patches, whose shape is $P \times P$, and flatten those patches into sequences $X_p \in \mathbb{R}^{ N \times 3P^2 }$. The position embedding is added to the patch embedding to retain positional information. A specialized class token $\mathbf{V}_{c l}^{(0)}$ is prepended to the embedded patches. Finally, the sequences of embedding vectors serve as input to the model as $\mathbf{h}^v \in \mathbb{R}^{ T_v \times d }$, where ${T_v}$ is the number of video feature vectors and ${d}$ is the dimension of each feature vector.

\subsection{Fusion via Efficient Video with Audio Encoding}
Video signals often contain high redundancy across multiple frames \cite{liu2022visually}, which may incur redundant computations. To address this issue and reduce the computational cost, we propose an efficient audio-visual fusion framework that incorporates temporal audio cues into static video frame representations using an audio-to-video attention mechanism. This method can be written as:

\begin{equation}
\mathbf{S}_t^{(\ell)}=\operatorname{MHA}\left(\mathbf{V}_t^{(\ell-1)}, \mathbf{V}_t^{(\ell-1)}, \mathbf{V}_t^{(\ell-1)}\right)+\mathbf{V}_t^{(\ell-1)}
\end{equation}

\begin{equation}
\mathbf{O}_t^{(\ell)}=\operatorname{MHCA}\left(\mathbf{S}_t^{(\ell-1)}, \mathbf{A}^{(\ell-1)}, \mathbf{A}^{(\ell-1)}\right)+\mathbf{S}_t^{(\ell-1)}
\end{equation}
where $\mathbf{V}_t^{(\ell-1)}$ is a visual patch representation, and $\mathbf{S}_t^{(\ell)} \in \mathbb{R}^{ (N+1) \times d}$ is the newly computed video representation, $\mathbf{A}^{(\ell-1)} \in \mathbb{R}^{ t \times d}$ is the audio representation at layer ${\ell-1}$, and $\mathbf{S}_t^{(\ell-1)} \in \mathbb{R}^{ (N+1) \times d }$ is the video representation at layer ${\ell-1}$. 
To calculate the new audio-visual representation $\mathbf{O}_t^{(\ell)}$, we utilize the MHCA mechanism, which performs a weighted summation of the audio-visual features. This allows the model to effectively incorporate long-range audio cues into the visual features. Due to the compact nature of the audio representation, this operation can be efficiently implemented.

We perform temporal pooling over the class tokens $\mathbf{AV}_{cl}^{(0)}$ across all video frames, resulting in the final audio-visual representation $\mathbf{f} \in \mathbb{R}^{d}$. To produce the final output embedding vector, three different learnable class tokens are employed. During training, the classification head is implemented using three MLPs with one hidden layer. This decoding strategy enables the model to produce multiple outputs as required, making it suitable for dense tasks like knowledge distillation or contrastive learning.

\subsection{Modality Dropout}
In multi-modal learning scenarios, it is common to encounter situations where one or more modalities may be absent or unavailable. 
To address this challenge, we employ modality dropout \cite{shi2022learning}, a technique designed to train the model to effectively handle missing modalities by randomly excluding specific modalities during training. This approach prevents the model from over-relying on a particular modality and thus enhances the overall robustness of the model by enabling the model to make predictions irrespective of the available modalities. In this work, we adopt modality dropout to prompt the model to learn how to optimally utilize the available modalities while handling instances where certain modalities are missing.

We employ modality dropout to mask the complete features of one modality before fusing the audio and visual inputs into the Transformer encoder. When both audio and video modalities are utilized as inputs, we assign a probability of $p_{av}$ for their joint selection. Conversely, when only one modality is used, the audio input is selected with a probability of $p_a$, while the video input is selected with a probability of $1-p_a$. The feature fusion process with modality dropout is mathematically represented as follows:
\begin{equation}
\label{modality dropout}
\mathbf{f}_t^{a v}= \begin{cases}\operatorname{concat}\left(\mathbf{f}_t^a, \mathbf{f}_t^v\right) & \text { with } p_{av} \\ \operatorname{concat}\left(\mathbf{f}_t^a, \mathbf{0}\right) & \text { with }\left(1-p_{av}\right) p_a \\ \operatorname{concat}\left(\mathbf{0}, \mathbf{f}_t^v\right) & \text { with }\left(1-p_{av}\right)\left(1-p_a\right)\end{cases}
\end{equation}
where concat means the channel-wise concatenation. We use the modality dropout at each iteration during the self-training. By incorporating modality dropout into our training procedure, we enable the U-FFIA model to effectively handle missing modalities and improve its robustness in real-world aquaculture applications.

\subsection{Unified Model with Knowledge Distillation}
Due to the limited number of frames used for both audio and visual inputs, the performance of the U-FFIA model can be compromised with the use of the single model architecture. To address this limitation, we employ knowledge distillation (KD) \cite{zhou2024rdnet} to enhance the performance of our audio and visual models.

Knowledge distillation \cite{hinton2015distilling} is a model compression technique in which a large-scale model (teacher) is compressed into a smaller model (student) while preserving similar performance. The key idea is to use the logits output by the teacher model (in our case, the pre-trained audio model CNN10 \cite{kong2020panns} and visual model S3D \cite{xie2018rethinking}) as the supervision information to guide the learning of the smaller student model. The teacher model uses all the audio and visual frames, while the student model uses a limited number of frames. During training, the teacher model is frozen, and the focus is solely on training the student model using the knowledge distillation loss.

We use the following loss for the student model training:
\begin{equation}
\begin{split}
\text{Loss} &= \lambda \cdot \text{Loss}_g(\psi(Z_s), y) \\
&\quad + (1-\lambda) \cdot \text{Loss}_d(\psi(Z_s), \psi(Z_t / \tau))
\end{split}
\end{equation}
where $\lambda$ is the balancing coefficient, $y$ is the ground truth of FFIA, $\operatorname{Loss}_g$ and $\operatorname{Loss}_d$ are the ground truth and distillation losses, respectively. We use cross-entropy (CE) and Kullback-Leibler divergence, for $\operatorname{Loss}_g$ and $\operatorname{Loss}_d$, respectively. $\psi$ is the activation function, $Z_s$ and $Z_t$ are the logits of the student and teacher model, respectively, and $\tau$ is the temperature parameter. For cross-model KD, the teacher and student may have different logit distributions, thus we only apply $\tau$ on the teacher logits to explicitly control the difference. 

\section{Dataset}
\label{sec:dataset}
To address the scarcity of publicly available datasets highlighted in the introduction, we introduce the AV-FFIA dataset, a comprehensive and large-scale dataset designed specifically for advancing research in multimodal FFIA. The AV-FFIA dataset aims to bridge the gap in the current literature by providing a diverse and well-annotated collection of synchronized fish feeding audio and video clips, enabling researchers to develop, compare, and evaluate novel multimodal FFIA methods. We will discuss the details in the following subsections.
\begin{figure}
\centering
\includegraphics[width=6.5cm]{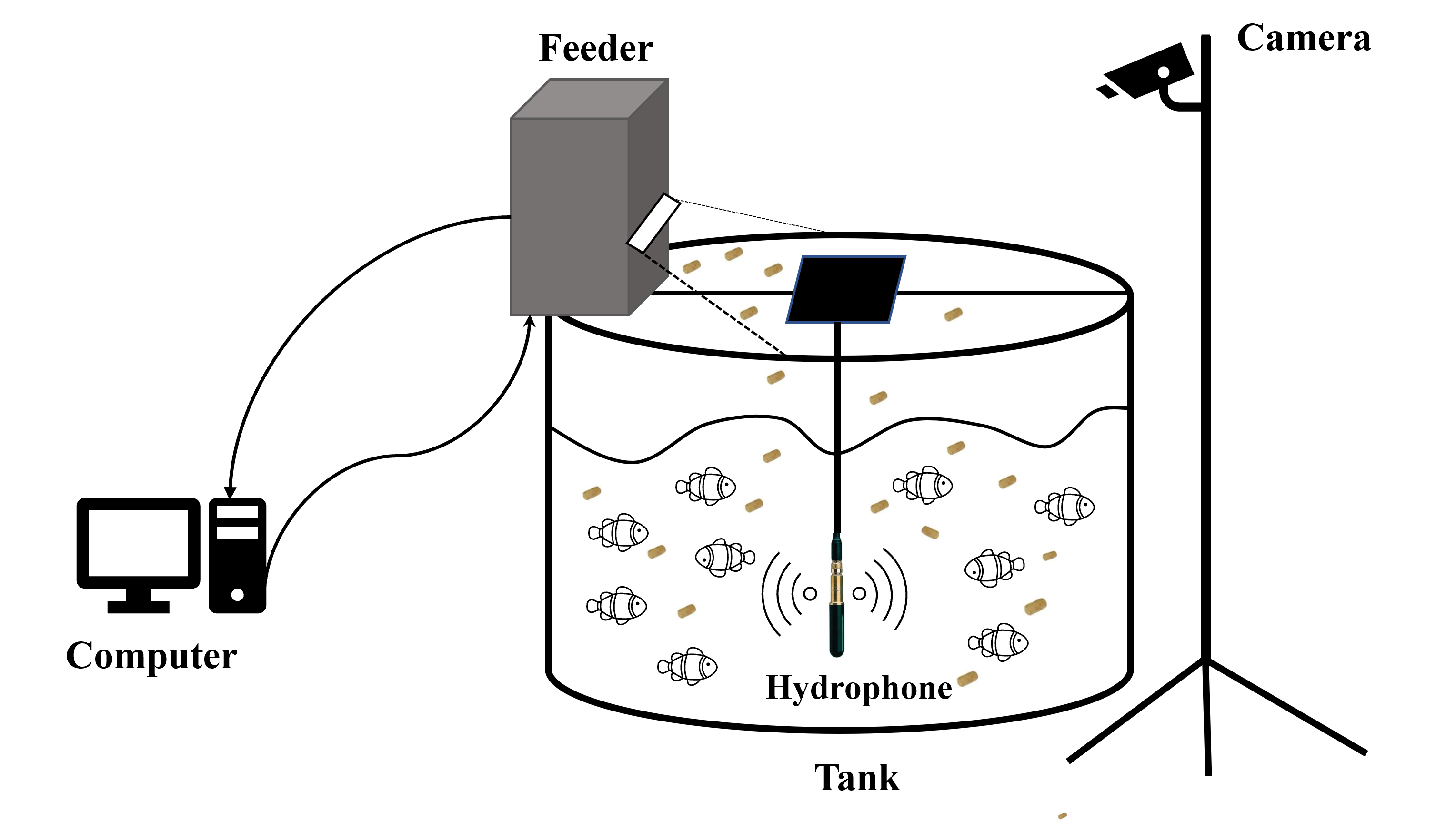}
\caption{Experimental systems for data collection. A hydrophone was underwater and the camera was deployed on a tripod with a height of about two meters to capture the video data.}
\label{fig:experiment system}
\end{figure}

\subsection{Challenges in Dataset Acquisition}
While acquiring the AV-FFIA dataset, we encountered several challenges that required careful consideration and planning.  Synchronizing audio and video recordings was a primary concern, which we addressed by employing a professional-grade recording system that ensured precise synchronization. Maintaining consistent lighting and water clarity was another challenge, as variations could impact data quality. We mitigated this issue by utilizing a controlled aquaculture environment with regulated lighting and filtration. Furthermore, capturing diverse fish feeding behaviours proved difficult due to the sporadic and unpredictable nature of fish feeding, necessitating lengthy recording sessions. To overcome this, we collaborated with experienced aquaculture technicians who provided guidance on optimal feeding schedules and techniques to elicit a wide range of feeding intensities. Despite these challenges, we successfully collected a comprehensive, well-annotated dataset that covers a broad spectrum of fish feeding behaviours, valuable for developing and evaluating multimodal FFIA methods.

\subsection{Dataset Collection}

The AV-FFIA dataset was collected in a controlled aquaculture environment using Oplegnathus Punctatus, a species of marine fish, as the experimental subject. The fish were farmed in a recirculating tank (\num{3} meters in diameter, \num{0.75} meters in depth) located in Yantai, Shandong Province, China. Each fish weighed approximately \num{150} grams.

To capture the audio and video data, we used a high-definition digital camera (Hikvision DS-2CD2T87E(D)WD-L) with a frame rate of \num{25} fps (\num{1920} × \num{1080}) and a high-frequency hydrophone (LST-DH01) with a sampling frequency of \num{256} kHz. The camera was deployed on a tripod at a height of about \num{2} meters to capture the video data, while the hydrophone was placed underwater to record the audio data (as shown in Fig. \ref{fig:experiment system}). The acquisition of video and audio data was synchronized to ensure the temporal alignment of the multimodal information.
 In the process of data collection, we followed the feeding rules in the real aquaculture production environment to ensure that the fish adapted to the laboratory environment as soon as possible and reduced the appetite loss caused by environmental changes. We feed the fish twice a day at \num{8} am and \num{5} pm.  The feeding process lasts about \num{3}-\num{15} minutes. Under the guidance of a fish feeding technician, we annotated the feeding video data as ``\textit{Strong}'',  ``\textit{Medium}'',  ``\textit{Weak}'' and ``\textit{None}'', based on the observed feeding intensity. To create a fine-grained dataset, we divided each one-minute video and audio clip into \num{30} segments with each segment in two seconds. Finally, 27,000 two-second video and audio clips are obtained, namely \textit{AV-FFIA}, with \num{6750} video and audio clips for each category of fish feeding intensity.
 The dataset was further split into training (21,000 clips), validation (2,800 clips), and testing (2,800 clips) sets by randomly selecting the audio clips and their corresponding video clips.

\begin{figure*}[b]
    \centering
    \includegraphics[scale=0.55]{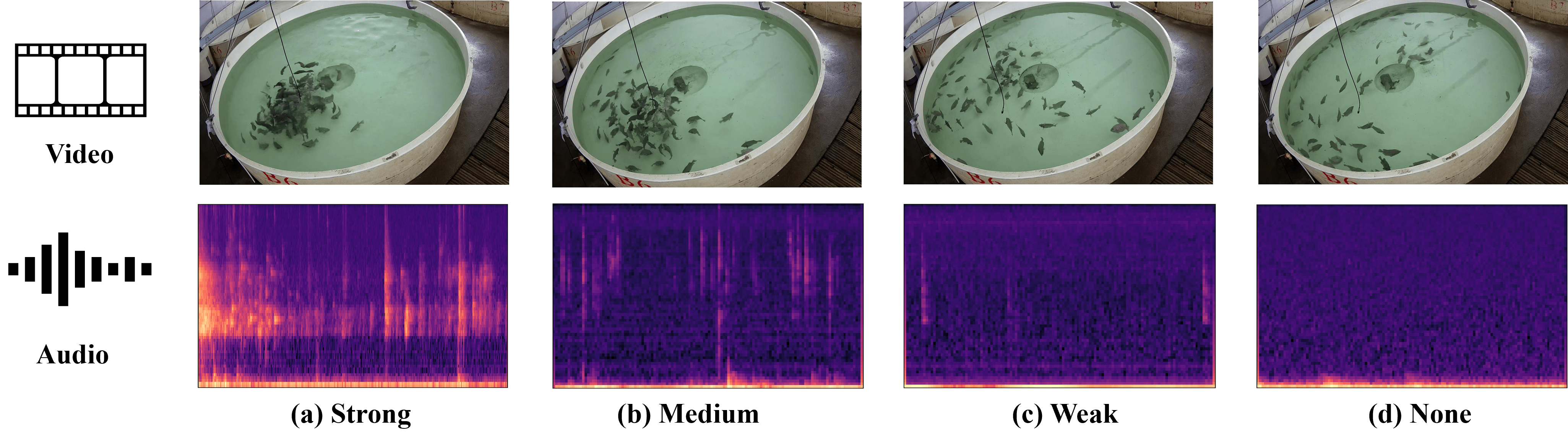}
    \caption{Video frames and mel spectrogram visualizations of four different fish feeding intensity: ``\textit{Strong}'', ``\textit{Medium}'', ``\textit{Weak}'' and ``\textit{None}''.}
    \label{fig:av-dataset}
\end{figure*}

\subsection{Dataset Visualization and Analysis}

To provide a better understanding of the AV-FFIA dataset, we randomly selected video frames and their corresponding audio mel-spectrograms for each feeding intensity class (Fig. \ref{fig:av-dataset}). The video frames show a clear relationship between the feeding intensity and the degree of fish aggregation, with higher intensity leading to denser fish grouping. Moreover, the continuous video footage reveals that the fish swimming speed and feeding activity increase with higher feeding intensity. The audio mel-spectrograms also exhibit distinct energy patterns corresponding to the fish feeding and chewing sounds. As the feeding process concludes, the energy spectrum gradually returns to normal levels, along with the decrease in fish aggregation and activity.

\subsection{Acoustic and Visual Noise}
Fish behaviour is influenced by various environmental factors, including natural ambient noise and visual disturbances. Incorporating audio-visual noise in fish behaviour analysis helps improve the robustness of the models. For the audio input corruption, we injected a bubble and pump noise to the entire audio with different signal-to-noise ratio (SNR) levels, -10 to 20 dB. For visual input corruption, there can be various noise types, additive noise, blur, colour distortion, occlusion, etc. We use the Gaussian noise, that often occurs in actual aquaculture. 
\section{Experiments}
\label{sec:experiment}
\subsection{Audio Data Preparation}
\subsubsection{Audio Data Processing}

\textcolor{black}{We use mel spectrogram as acoustic features, which has been widely used for audio classification \cite{kong2020panns}.} The original audio sampling rate is \num{256} kHz, which can lead to computational and storage burdens. We down-sample the audio signals to \num{64} kHz to reduce the computational complexity. Then, we calculate the short-time Fourier transform (STFT) with a Hanning window of \num{2048} samples and a hop size of \num{1024} samples, and finally, we apply the mel filter banks with \num{128} bins. Therefore, for a \num{2}-second audio signal, we have a mel spectrogram with the shape of \num{128} $\times$ \num{128}. We use the SimPF method \cite{liu2023simple} which is a simple non-parametric pooling operation, to reduce the redundant information within the mel-spectrogram, leading to a shape of \num{64} $\times$ \num{128}  mel spectrogram. This streamlined mel spectrogram representation enhances the efficacy of audio classification tasks. To evaluate the noise robustness of the model, we add bubble and pump noise to the entire audio signal with different SNR levels, ranging from -\num{10} dB to \num{20} dB. This allows us to assess the model's performance under varying noise conditions.

\subsubsection{Loss and Training}
During training, we use the SpecAugment method \cite{park2019specaugment} to expand our training samples. With SpecAugment, the spectrogram can be modified by masking blocks of consecutive frequency channels and masking blocks of time steps. We employ the cross-entropy loss function to train our audio classification model. The cross-entropy loss quantifies the dissimilarity between the predicted class probabilities and the ground truth class labels. By minimizing this loss, we effectively guide the model to learn the correct class boundaries and improve its classification performance.

\subsection{Video Data Processing}
The original video comprises \num{50} frames, each with a resolution of \num{2560}$\times$\num{1440}. To reduce computational complexity and redundancy across multiple frames, we randomly select \num{4} frames from the video and resize each frame to a smaller dimension of \num{224}$\times$\num{224}. This resizing step aims to retain essential visual information while making the data more manageable for training purposes. By combining downsampling, random cropping, and colour augmentation techniques during training, we effectively utilize the high-resolution video data while promoting the model's ability to generalize well to diverse inputs. To evaluate the model's robustness to visual noise, we corrupt the video frames by adding darkness and Gaussian noise. The darkness is applied by reducing the pixel intensities, while the Gaussian noise is added with a maximum variance of \num{0.2}. These noise corruptions help to assess the model's performance under challenging visual conditions.

\subsection{Benchmarking Methods}
To evaluate the performance of various approaches on the AV-FFIA dataset and establish baselines for future research, we conducted benchmark studies on audio and visual modalities.

\subsubsection{Audio-Based Models}
For audio-based FFIA, we employed the lightweight MobileNetV2 \cite{sandler2018mobilenetv2} architecture as the backbone for our models to balance performance and computational efficiency in aquaculture applications. We also use the pre-trained PANNs \cite{kong2020panns} and AST \cite{gong2021ast} models on our AV-FFIA dataset to leverage their strong performance on audio classification tasks and investigate their effectiveness for FFIA.

\subsubsection{Video-Based Models}

In the realm of video-based FFIA, we explored the performance of state-of-the-art 3D CNN models, including I3D \cite{carreira2017quo} and 3D-ResNet \cite{dubey20193d}, which have shown promising results in various video tasks. Additionally, we investigated the potential of Transformer-based video classification models, such as ViViT \cite{arnab2021vivit} and 3D-ViT \cite{bertasius2021space}, on our AV-FFIA dataset, as these models have demonstrated state-of-the-art results on benchmark datasets like Kinetics-400 \cite{kay2017kinetics} and Kinetics-600 \cite{carreira2018short}. To balance performance and model size, we also employed the Separable 3D CNN (S3D) \cite{xie2018rethinking} model, which replaces the original I3D \cite{carreira2017quo} convolutions with spatial and temporal-separable 3D convolutions, resulting in improved performance and reduced computational complexity compared to the I3D \cite{carreira2017quo} model.

\subsection{Experimental Setups}
All the models are trained using the Adam optimizer \cite{kingma2014adam}, and the training and evaluation are performed on a Nvidia-RTX-3090Ti-24GB GPU.

\subsubsection{Audio Experiment Setup}

For the audio-based FFIA experiments, we consider two baseline models: the model trained from scratch and the pre-trained model on AudioSet \cite{gemmeke2017audio} and fine-tuned on our dataset. We train the following models on the AV-FFIA dataset from scratch, such as CNN6,  CNN10, CNN14, ResNet18, ResNet22, MobileNetV1, and MobileNetV2 and those models use the same structure as that in PANNs \footnote{ \url{https://github.com/qiuqiangkong/audioset_tagging_cnn}} \cite{kong2020panns}. We train these models with a batch size of 200 and 100 epochs, using a learning rate 0.001. For the fine-tuning approaches, we leverage pre-trained models, such as CNN6, CNN10, MobileNetV2 and AST \footnote{ \url{https://github.com/YuanGongND/ast}}\cite{gong2021ast}. We fine-tune these pre-trained models on the AV-FFIA dataset with a batch size of 200 and 20 epochs. We use a learning rate of 0.001 for the CNN models, and 0.0001 for AST, respectively.

\subsubsection{Video Experiment Setup}

For the video-based FFIA experiments, we also consider two baseline models: the model trained from scratch and the pre-trained model on Kinetics-600 \cite{carreira2018short}  and fine-tuned on our dataset. We train the following models on the AV-FFIA dataset from scratch, such as I3D \cite{carreira2017quo}, 3D-ResNet10 \cite{wang2021deep}, 3D-ResNet18 \cite{ebrahimi2020introducing}, S3D \cite{xie2018rethinking}. The code can be found at PyTorchConv3D \footnote{ \url{https://github.com/tomrunia/PyTorchConv3D}}. We train these models with a batch size of 40, 100 epochs, using a learning rate 0.001. For the fine-tuning approaches, we leverage the pre-trained models, such as 3D-ViT \footnote{ \url{https://github.com/lucidrains/vit-pytorch/tree/main}} \cite{bertasius2021space}, ViViT \footnote{ \url{https://github.com/rishikksh20/ViViT-pytorch}} \cite{arnab2021vivit} and S3D \cite{xie2018rethinking}. We fine-tune these pre-trained models on the AV-FFIA dataset with a batch size of 40 and 20 epochs. We use a learning rate of 0.001 for the S3D \cite{xie2018rethinking} models, and 0.001 for the 3D-ViT \cite{bertasius2021space} and ViViT \cite{arnab2021vivit} models, respectively. 

\subsubsection{Audio-Visual Fusion Setup}
 We use the ViT architecture \cite{dosovitskiy2020image} ($L =$ \num{6}, $N_H =$ \num{8}, $d = $\num{1024}) as our backbone.  
The model is trained with a batch size of \num{20}, the number of epochs \num{200}, and the Adam optimizer with a learning rate of \num{0.0001}.

\subsubsection{Unified Mixed-Model Setup}
 For the unified mixed-model experiments, we also use the same ViT architecture as our backbone. We use Adam optimizer with a learning rate of \num{0.0001} for training the model. The batch size is set to \num{20} and the number of epochs is \num{200}. For knowledge distillation during the training, we fix the hyperparameters $\lambda=0.5$ and $\tau=2.5$, which control the balance between the teacher and student models' contributions and the temperature of the softened probabilities, respectively.

\subsection{Evaluation Metrics}

Accuracy is the primary evaluation metric used in our experiments, as it is widely adopted in the literature on FFIA classification \cite{cui2022fish, feng2022fish, zhou2019evaluation}. Accuracy refers to the ratio of correct predictions to the total number of predictions, providing a straightforward and intuitive measure of the overall performance of a classification model. By using accuracy as the performance metric, we can directly compare our results with previous methods in the literature.

\section{Results and discussion}
\label{sec:results}
\subsection{The Benchmark Results of Audio-Based FFIA}

\begin{table}
\caption{The results of the different audio-based models on AV-FFIA. The FLOPs are computed for one 2-second audio clip with a 64 kHz sampling rate.}
\centering
\label{tab: audio result}
\begin{tabular}{cccc}
\hline
Model                                                                & Accuracy  & Parameters (M) & FLOPs\\ \hline
MobileNetV1 \cite{howard2017mobilenets}                              & 0.823   & 4.3   &198.275 \\ 
MobileNetV2 \cite{sandler2018mobilenetv2}                            & 0.824   & 3.5  & \textbf{118.553M}  \\ 
MobileViT   \cite{mehta2021mobilevit}                                & 0.794   & 2.3  &229.910M \\ 
ResNet18    \cite{yu2019abnormality}                                 & 0.812   & 39.5  &2.676G \\ 
ResNet22    \cite{verbitskiy2021eranns}                              & 0.818   & 62.6  &3.605G \\ 
CNN6        \cite{kong2020panns}                                     & 0.823   & 4.6   &2.551G \\ 
CNN10       \cite{kong2020panns}                                     & 0.832   & 5.0  &3.347G \\ 
CNN14       \cite{kong2020panns}                                     & 0.856   & 79.7  &5.164G \\ 
Pre-AST     \cite{gong2021ast}                                       & 0.736   & 85.3  &12.440G \\
Pre-CNN6    \cite{kong2020panns}                                     & 0.847   & 4.6 &2.551G   \\ 
Pre-CNN10   \cite{kong2020panns}                                     & \textbf{0.859}   & 5.0   &3.347G \\ 
\textbf{Pre-MobileNetV2}  \cite{kong2020panns}                     & \textbf{0.857}   & \textbf{1.9} & \textbf{133.958M}   \\ \hline
\end{tabular}
\end{table}

\begin{figure}[htbp]
\centering
\begin{minipage}[t]{0.48\textwidth}
\includegraphics[width=8.5cm]{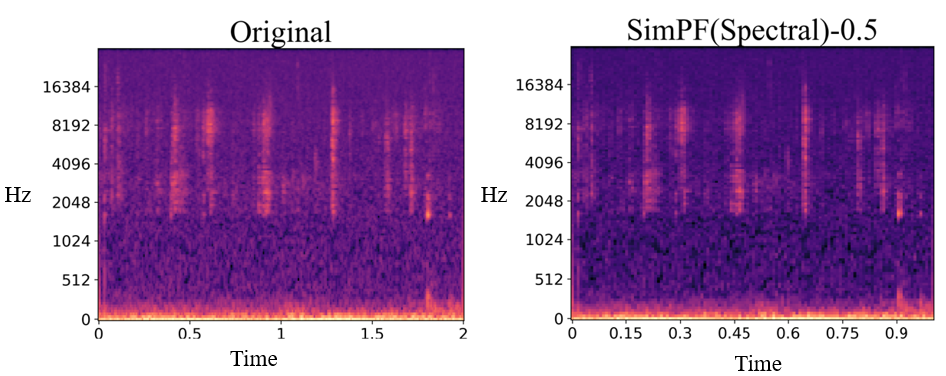}
\caption{Visualization of the impact of SimPFs on the mel-spectrogram of a FFIA audio clip with 50\% compression factor.}
\label{fig:simpf}
\end{minipage}
\end{figure}

\begin{table}[htbp]
\caption{The results of the different video-based models on the AV-FFIA dataset. The FLOPs are computed for one 2-second video clip with 20 frames randomly chosen.}
\label{tab:video results}
\centering
\begin{tabular}{ccccc}
\hline
Models                                                    & Accuracy  & Parameters (M)   & FLOPs (G) \\ \hline
I3D  \cite{carreira2017quo}                               & 0.791   & 12.3    & 35.054     \\ 
3D-ResNet10  \cite{wang2021deep}                          & 0.855   & 14.4    & 28.602     \\ 
3D-ResNet18   \cite{ebrahimi2020introducing}              & 0.874   & 33.2    & 42.219     \\ 
Pre-3D-ViT   \cite{bertasius2021space}                        & 0.882   & 27.8    & 77.614     \\ 
Pre-ViViT     \cite{arnab2021vivit}                           & 0.743   & 78.2    & 77.344     \\ 
S3D   \cite{xie2018rethinking}                            & 0.816   & 7.9     & 22.826     \\ 
\textbf{Pretrained-S3D}                                   & \textbf{0.898}   & \textbf{7.9}    & \textbf{22.826}     \\
\hline
\end{tabular}
\end{table}


\begin{table*}[]
\caption{The Performance of different AV fusion methods on the AV-FFIA dataset and AV fusion under noisy environment.}
\label{tab:AV result}
\centering
\begin{tabular}{ccccccccccc}
\hline Model & A only & V only & AV Fusion   & $-10 \mathrm{~dB}$ & $-5 \mathrm{~dB}$ & $0 \mathrm{~dB}$ & $10 \mathrm{~dB}$ & $20 \mathrm{~dB}$ \\ \hline
 Self-attention \cite{vaswani2017attention}  & 0.821 & \textbf{0.883} & 0.892   & 0.713 & 0.751 & 0.774 & 0.826 & 0.863 \\
 Cross-attention \cite{praveen2022joint}     & n/a & n/a & \textbf{0.917}   & \textbf{0.727} & 0.768 & 0.805 & 0.834 & 0.867 \\
 MBT \cite{nagrani2021attention}             & n/a & n/a & 0.724  & 0.617   & 0.631 & 0.652 & 0.681 & 0.713 \\
U-FFIA                                       & 0.786 & 0.836 & 0.892   & 0.721 & 0.763 & 0.797 & 0.836 & 0.872 \\
\textbf{U-FFIA + KD (Proposed)}              & \textbf{0.824} & 0.857 & 0.907  & 0.726 & \textbf{0.779} & \textbf{0.806} & \textbf{0.835} & \textbf{0.876} \\
\hline
\end{tabular}
\end{table*}

\begin{table*}[]
\caption{The FLOPs and parameters of different AV fusion methods on the AV-FFIA dataset.}
\label{tab: AV_parameters}
\centering
\begin{tabular}{lllllll}
\hline
\multirow{2}{*}{Model} & \multicolumn{3}{c}{FLOPs} & \multicolumn{3}{c}{Parameters} \\ \cline{2-7} 
                                            & A (M)     & V  (G)     & AV (G)     & A (M)       & V  (M)      & AV (M)      \\ \hline
Self-attention \cite{vaswani2017attention}  & 210.346  & 68.193      & 68.404 & 39.781        & 45.710        & 47.695        \\ 
Cross-attention  \cite{praveen2022joint}    & n/a      & n/a         & 60.933       & n/a        & n/a        & 35.082        \\ 
MBT      \cite{nagrani2021attention}        & n/a      & n/a         & 68.690       & n/a        & n/a        & 47.689        \\ 
\textbf{U-FFIA}                             & \textbf{105.336}        & \textbf{15.401}      & \textbf{15.488 }      & \textbf{21.032}        & \textbf{19.506}        & \textbf{21.626}        \\ \hline
\end{tabular}
\end{table*}

\begin{table*}[]
\centering
\caption{Ablation study for modality dropout probabilities on U-FFIA.}
\label{tab:ablation}
\begin{tabular}{ccc|c|c|ccccc}
\hline \multicolumn{3}{c|}{ PT } & \multicolumn{1}{c|}{ Modality } & \multicolumn{1}{c|}{ Clean } & \multicolumn{5}{c}{ Noise} \\
\hline ${p_{av}}$ & ${p_a}$ & ${p_v}$ & &  & $-10 \mathrm{~dB}$ & $-5 \mathrm{~dB}$ & $0 \mathrm{~dB}$ & $10 \mathrm{~dB}$ & $20 \mathrm{~dB}$ \\ \hline
 1 & 0 & 0 & AV & \textbf{0.917}  & 0.798 & 0.819 & 0.826 & 0.864 & 0.887 \\
 0.7 & 0.15 & 0.15 & AV & 0.912  & 0.781 & 0.803 & 0.812 & 0.846 & 0.881 \\
 0.5 & 0.25 & 0.25 & AV & 0.895  & 0.726 & 0.779 & 0.806 & 0.832 & 0.872 \\
 0.4 & 0.2 & 0.4 & AV & 0.862  & 0.702 & 0.768 & 0.776 & 0.837 & 0.853 \\ \hline
 0.7 & 0.15 & 0.15 & A & 0.714  & 0.581 & 0.602 & 0.632 & 0.656 & 0.682 \\
 0.5 & 0.25 & 0.25 & A & \textbf{0.786}  & 0.596 & 0.625 & 0.654 & 0.698 & 0.732 \\
 0.4 & 0.2 & 0.4 & A & 0.761  & 0.588 & 0.612 & 0.646 & 0.671 & 0.703 \\ \hline
 0.7 & 0.15 & 0.15 & V & 0.740  & 0.684 & 0.684  & 0.684  & 0.684  & 0.684  \\
 0.5 & 0.25 & 0.25 & V & 0.836  & 0.788 & 0.788 & 0.788 & 0.788 & 0.788 \\
 0.4 & 0.2 & 0.4 & V & \textbf{0.858}  & 0.803 & 0.803 & 0.803 & 0.803 & 0.803 \\
\hline
\end{tabular}
\end{table*}

We conducted an extensive evaluation of various common methods for FFIA on our AV-FFIA dataset, including CNN \cite{hu2012fish}, ResNet \cite{targ2016resnet}, and MobileNet \cite{howard2017mobilenets} models.  Table \ref{tab: audio result} presents the performance of these baseline models, revealing that the pretrained-CNN10 \cite{kong2020panns} and pretrained-MobileNetV2 \cite{kong2020panns} models achieved remarkable performance, giving an accuracy higher than \num{0.85}.
To gain insights into the computational complexity of these baseline models, we analyzed their parameters. Surprisingly, the pretrained-MobileNetV2 \cite{kong2020panns} model, with a modest \num{1.9} million parameters, delivered competitive results, outperforming larger CNN models while maintaining a more compact size. 

The results also indicated that pre-trained models PANNs \cite{kong2020panns} and AST \cite{gong2021ast} outperformed the models trained from scratch, i.e. MobileNetV1 \cite{howard2017mobilenets}, ResNet18 \cite{yu2019abnormality}  highlighting the benefits of transfer learning for the FFIA task. \textcolor{black}{The transformer models, such as \cite{vaswani2017attention}, which were originally designed for sequential data in natural language processing (NLP) tasks, are built on the self-attention mechanisms.} While these mechanisms excel in modelling long-range dependencies, they might struggle to discern fine-grained local patterns evident in audio spectrograms, leading to suboptimal performance on the AV-FFIA dataset compared to CNN-based models. 

We also investigated the effectiveness of the SimPF method with varying compression coefficient settings for the MobileNetV2 models.  As shown in Fig. \ref{fig:simpf}, even with a \num{50}\% compression setting, the classification accuracy decreased only marginally by \num{0.01}\%, while the number of FLOPs was reduced by half to \num{66.979} M. This demonstrates that the SimPF method can effectively reduce redundant information in the mel-spectrogram without significant loss in performance, making it a promising technique for improving computational efficiency in FFIA applications.

\subsection{The Benchmark Results of Video-Based FFIA}

We compared commonly used video classification models, including I3D \cite{carreira2017quo}, S3D \cite{xie2018rethinking}, and 3D-ViT \cite{bertasius2021space} models, on AV-FFIA dataset. Table \ref{tab:video results} presents the performance of different video-based baseline models on the AV-FFIA dataset, with S3D \cite{xie2018rethinking} models achieving the highest accuracy of \num{0.898}, outperforming other models. Compared with the performance of the S3D \cite{xie2018rethinking} models, the performance of 3D-ViT \cite{bertasius2021space} and ViViT \cite{arnab2021vivit} models reached approximately \num{0.88} and \num{0.74}, respectively. This discrepancy can be attributed to several factors. Firstly, ViT-based architectures may have limitations in capturing temporal information, which is crucial for video-based FFIA. Secondly, the relatively smaller size of the AV-FFIA dataset compared to large-scale video datasets like Kinetics \cite{carreira2018short} might lead to overfitting in more complex models like 3D-ViT \cite{bertasius2021space} and ViViT \cite{arnab2021vivit}. Lastly, differences in model complexity and parameter efficiency could also contribute to the performance gap. S3D \cite{xie2018rethinking} models are specifically designed to capture spatiotemporal information in videos using 3D convolutions. Additionally, they are often pre-trained on large-scale video datasets like Kinetics, which encompass diverse and representative video clips. This pre-training allows S3D models to learn robust and generalizable features that can be effectively transferred to the FFIA task through fine-tuning. One notable advantage of S3D \cite{xie2018rethinking} models is their ability to strike a balance between computational complexity and model performance. As evident from their fewer parameters (\num{7.9} million) and FLOPs (\num{22.826} G) compared to I3D \cite{carreira2017quo} and 3D-ResNet \cite{hara2017learning} models, they are more efficient, making them a promising choice for applications in aquaculture.

\subsection{The Benchmark Results of Audio-Visual Based FFIA}

We compared various audio-visual fusion methods, such as cross-attention \cite{praveen2022joint}, self-attention \cite{vaswani2017attention}, and multimodal bottleneck transformer (MBT) \cite{nagrani2021attention} on the AV-FFIA dataset. In Table \ref{tab:AV result}, we present the results of different baseline methods based on audio-visual modalities on AV-FFIA. Compared with the performance achieved with individual modalities, we observed that all the audio-visual fusion methods surpassed any single modality, which indicates that the multimodal approach significantly outperforms the single-modality based approaches. In addition to the improved accuracy, we found that the audio-visual approach outperforms the single modality based approaches under noisy environments. This finding highlights the importance of leveraging complementary information from multiple modalities to enhance the model's resilience to noise commonly encountered in aquaculture settings. Among the audio-visual fusion methods, the self-attention method achieved an accuracy of \num{0.892}, while the cross-attention method achieved \num{0.917}. We believe that the superiority of the cross-attention method can be attributed to its ability to selectively focus on relevant information in both audio and visual modalities. In contrast, the self-attention method treats all information equally, which may not be optimal for capturing the complex interactions between modalities.

To illustrate this point, we can consider the fish feeding scenario, \textcolor{black}{where the fish may quickly consume all the current feed, resulting in no sound being produced.} However, the video still shows the fish highly aggregated and in a state of hunger. In such cases, the cross-attention method allows the information from the audio modality to complement that from the visual modality, enabling the model to make more accurate predictions. On the other hand, the MBT \cite{nagrani2021attention} method achieved a lower accuracy of only \num{0.724} compared to the cross-attention method. This performance gap could be attributed to attention bottlenecks that attend to each modality sequentially, leading to a failure to capture the complex relationships among modalities. Consequently, there may be information loss and a lack of direct interaction between modalities, which can limit the model's ability to effectively fuse the audio and visual information.

\subsection{The Results of Unified Model Based FFIA}

Table \ref{tab:AV result} provides an in-depth analysis of our proposed model and several audio-visual fusion methods that are currently in use. Our U-FFIA model achieved an accuracy of \num{0.786} and \num{0.836} for audio and video single modalities, respectively, and an accuracy of \num{0.907} for audio-visual fusion. It's worth noting that the use of fewer audio and video frames can slightly degrade the performance. However, knowledge distillation can significantly improve the performance of U-FFIA across both audio and video modes. Through the application of the knowledge distillation approach, we observed marked improvements in both audio and visual performance (\num{0.824} and \num{0.857}, respectively). These enhancements place the individual modality based model on par with other audio-visual fusion methods, including self-attention and cross-attention, in terms of performance. Furthermore, we conducted a thorough analysis of computational complexity and model parameters across various models (refer to Table \ref{tab: AV_parameters}). In comparison to a Transformer model employing all video frames, our proposed model with fewer video frames demonstrates a remarkable reduction of \num{57}\% in parameters, accompanied by a substantial \num{77}\% decrease in FLOPs, while maintaining a negligible \num{3}\% dip in performance. This efficiency is mirrored in the audio domain, where our model showcases a \num{48}\% decrease in model size in terms of the number of parameters and a \num{50}\% reduction in FLOPs, but with a marginal performance variance. Importantly, our proposed unified model exhibits comparable performance on audio-visual fusion while concurrently reducing the number of parameters and FLOPs by \num{54}\% and \num{77}\%, respectively. This compelling evidence underscores the efficiency of our proposed model. Additionally, our proposed model has better robustness, as shown in Table \ref{tab:AV result}. The proposed model demonstrate promising performance in AV-FFIA in both effectiveness and robustness.



\subsection{Ablation Studies}

To gain a deeper understanding of the impact of modality dropout configurations on our AV-FFIA dataset during training, we consider four different modality dropout configurations. These configurations are represented by the probabilities of using both audio and video streams $p_{av}$, only the audio stream $p_{a}$, and only the video stream $p_{v}$, as shown in Table \ref{tab:ablation}.
The first configuration, denoted as ($p_{av}$, $p_{a}$, $p_{v}$) = (1.00, 0.00, 0.00), corresponds to not employing any modality dropout. When training the U-FFIA model, we observed that increasing the value of $p_{av}$ in models trained with modality dropout resulted in slightly worse performance on the visual-only and audio-only test sets. We attribute this outcome to an imbalance in the training data, leading to overfitting in the audio-visual modality but insufficient training in the single modalities.

In contrast, a slight increase in the probability within the video modality resulted in a slight improvement in video performance. This improvement can be attributed to the simple linear transformation undergone by the video modality's input, capturing shallow visual features. Furthermore, to reduce computational complexity, we randomly down-sampled the video, resulting in a reduced number of video frames. Consequently, achieving optimal performance may necessitate additional training iterations.
For the audio modality, the high-dimensional features are extracted from pre-trained models, as a result, satisfactory performance could be achieved with fewer iterations. Nevertheless, all three modality configurations significantly outperformed models trained without modality dropout, showing the importance of modality dropout in training a single model for three distinct modalities.  This result demonstrates that modality dropout is an effective technique for improving the U-FFIA model's ability to handle missing modalities and enhance its robustness in real-world scenarios.

\begin{figure}[hbt]
\centering
\includegraphics[width=8.5cm]{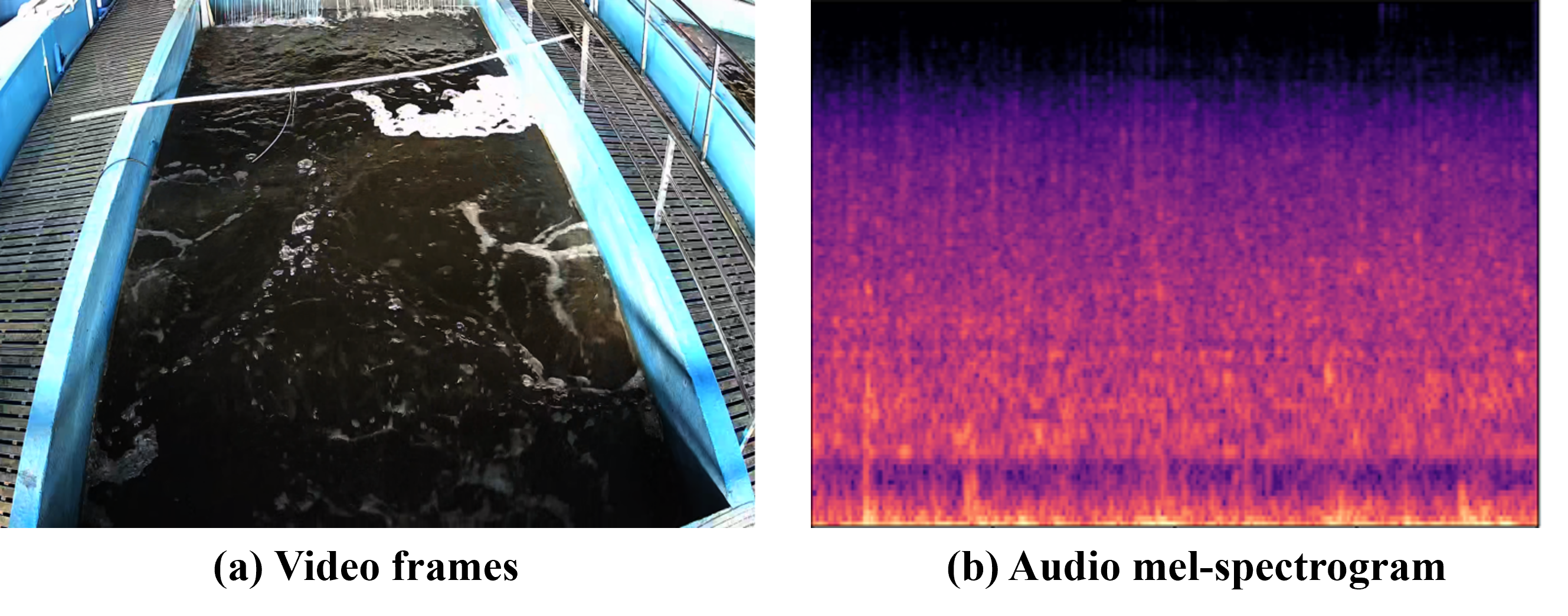}
\caption{Real aquaculture environment for Tilapia rearing. The turbidity of the water and low lighting conditions make it challenging to visually observe the fish. (a) shows a video frame during fish feeding, and (b) shows the mel-spectrogram of the corresponding audio clip.}
\label{fig: real environment}
\end{figure}

\subsection{Experiments in Real Aquaculture Environment}
\label{sec: real data experiment}

\textcolor{black}{To demonstrate the real-world applicability of our method, we conducted experiments with a commercial aquaculture facility at the Guangzhou Aquatic Products Promotion Station, China (as shown in Fig. \ref{fig: real environment}). We collected 12,700 audio-visual samples of Tilapia fish, a common aquaculture species, from two identical tanks (\num{4}m x \num{2}m x \num{3}m). This real-world environment presented challenges such as environmental noise, water surface reflection, and foams, which are not present in controlled settings. The dataset maintained the same specifications as our controlled dataset, with \num{2}-second audio-visual clips annotated by experienced technicians. We randomly split the real-world dataset into training (\num{70}\%), validation (\num{10}\%), and testing (\num{20}\%) sets.}

\textcolor{black}{To adapt our model for this real data, we employed a two-stage training approach. First, we pretrained our model on the AV-FFIA dataset and then fine-tuned it on real-world data. This approach allowed us to leverage the large-scale AV-FFIA dataset while adapting the model to the specific conditions of the real aquaculture environment. Our fine-tuned U-FFIA model achieved \num{0.892} accuracy on the real-world dataset, outperforming the SOTA models (e.g., cross-attention \cite{praveen2022joint}, self-attention \cite{vaswani2017attention}, and MBT \cite{nagrani2021attention}). We observed that on the real dataset, the video performance (\num{0.822}) decreased significantly due to the darkness and overlap of the foams. However, the audio performance (\num{0.816}) did not decrease as much because the pump and water flow operate at stable frequencies, which are quite different from the frequencies of fish slapping water or ingesting feed. This frequency difference allows the audio modality to maintain its performance even in the presence of background noise.}

The results demonstrate that our proposed method maintains a high level of performance in the real aquaculture environment, with only a slight decrease in accuracy compared to the controlled experiments. The successful application of our proposed method in a commercial aquaculture facility highlights its practicality and potential for real-world deployment. Despite the challenges posed by the uncontrolled environment, such as variations in water quality, background noise, and lighting conditions, our method demonstrates strong performance and robustness.

\section{Conclusion and Future Work}
\label{sec:conclusion}

In this paper, we introduced U-FFIA, a novel unified mixed-modality based method for fish feeding intensity assessment, capable of processing audio, visual, or audio-visual modalities. We also presented AV-FFIA, a large-scale audio-visual dataset for FFIA research. This dataset fills a crucial gap in the field, providing a comprehensive resource for developing and evaluating multimodal FFIA methods. Through extensive benchmarking experiments on the AV-FFIA dataset, we demonstrated the superior performance of the U-FFIA model compared to state-of-the-art modality-specific FFIA models, especially under noisy environments. In addition to the controlled experiments, we validated the performance of our proposed method in real-world scenarios through experiments conducted in a commercial aquaculture facility. The results demonstrate that our method maintains a high level of performance in the real aquaculture environment, with only a slight decrease in accuracy compared to the controlled experiments.
Moreover, the U-FFIA model achieves this level of performance while requiring significantly fewer computational resources as compared with baseline methods, making it an efficient solution for FFIA. In future work, we plan to expand the AV-FFIA dataset by including additional fish species and focus on developing on-device models tailored to the specific needs of aquaculture applications.
\section{ACKNOWLEDGMENT}
This work was supported by the National Natural Science Foundation of China ``Dynamic regulation mechanism of nitrogen in industrial aquaponics under the condition of asynchronous life cycle'' [Grant no. 32373186], Digital Fishery Cross-Innovative Talent Training Program of the China Scholarship Council (DF-Project) and a Research Scholarship from the China Scholarship Council. Ethics approval for this study was obtained from the Welfare and Ethical Committee of China Agricultural University (Ref: AW30901202-5-1). For the purpose of open access, the authors have applied a Creative Commons Attribution (CC BY) licence to any author-accepted manuscript version arising.
\begin{bibliography}{mybib}
\bibliographystyle{ieeetr}
\end{bibliography}
\vspace{-30pt}
\begin{IEEEbiography}[{\includegraphics[width=1in,height=1.25in,clip,keepaspectratio]{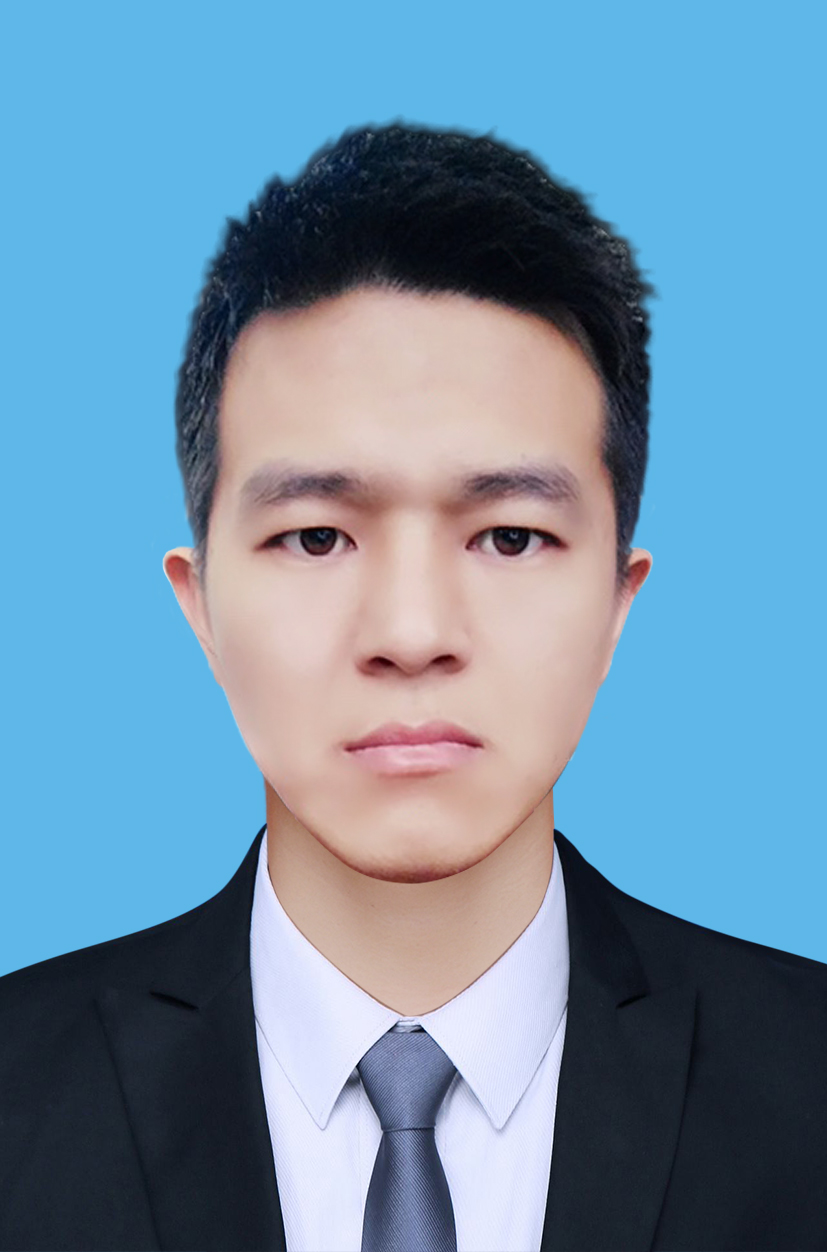}}]{Meng Cui}
received the M.E. degree from the HeBei Agriculture University, Baoding, China, in 2019. He is currently working toward a Ph.D. degree from the Centre for Vision, Speech and Signal Processing (CVSSP) at the University of Surrey. His
research topic includes audio signal processing and multimodal machine learning.
His current research is mainly on multimodal machine learning for fish behaviour analysis, especially fish feeding behaviour analysis.
\end{IEEEbiography}
\vspace{-30pt}

\begin{IEEEbiography}[{\includegraphics[width=1in,height=1.25in,clip,keepaspectratio]{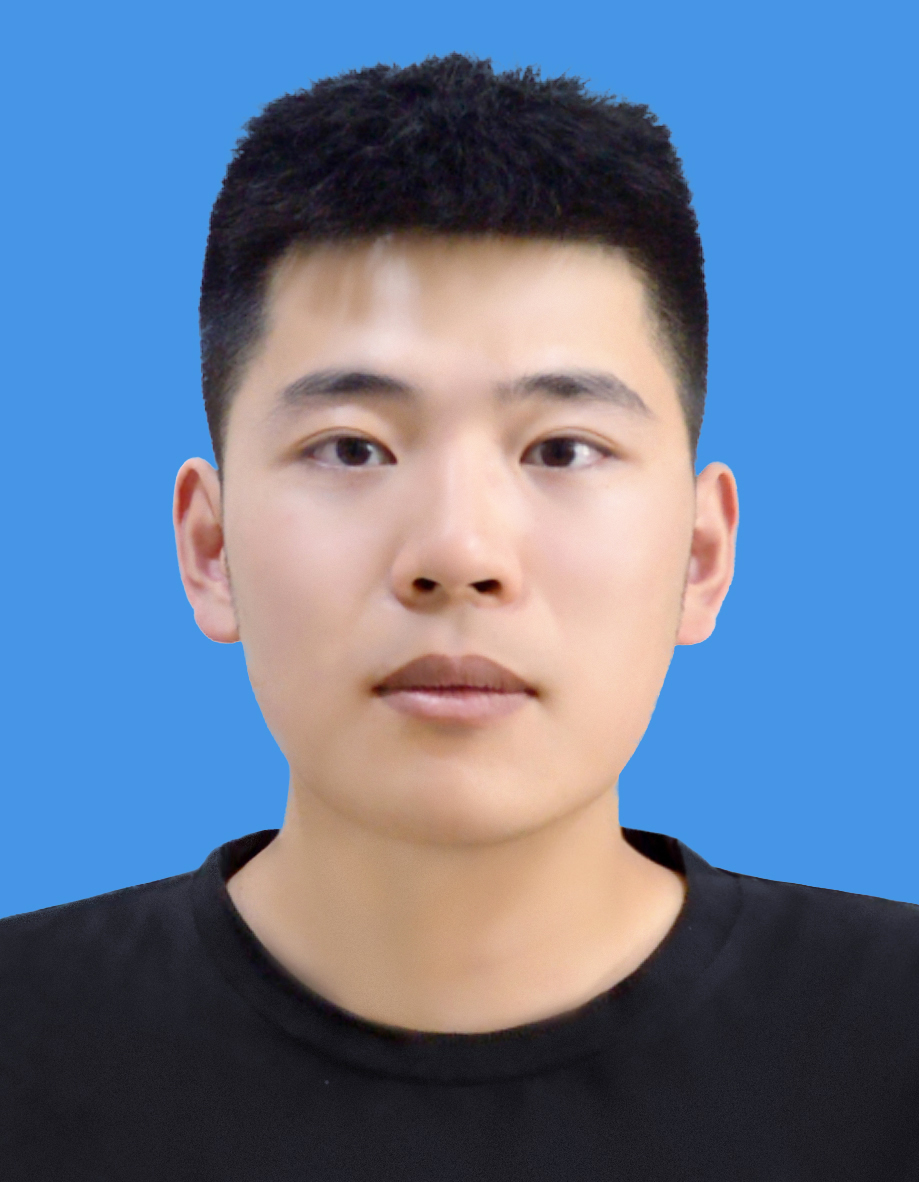}}]{Xubo Liu}
received the B.Eng. degree in Telecommunications Engineering from the Queen Mary University of London, in
2020. He is currently working toward a Ph.D. degree from the Centre for Vision, Speech and Signal Processing (CVSSP) at the University of Surrey. As a PhD student, he has worked on a number of
projects in the area of audio separation and text-to-sound generation using machine learning methods. His research topic is cross-modal translations between audio and texts, including text-to-sound generation and automated audio captioning.
\end{IEEEbiography}
\vspace{-30pt}
\begin{IEEEbiography}[{\includegraphics[width=1in,height=1.25in,clip,keepaspectratio]{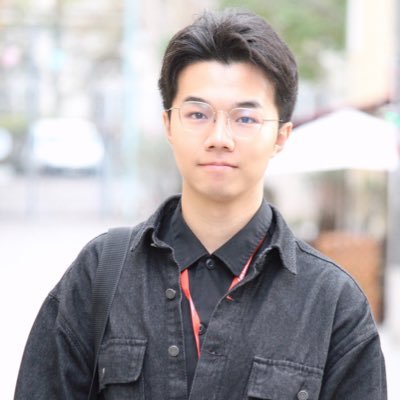}}]{Haohe Liu}
is a Ph.D. student at the Centre for Vision Speech and Signal Processing (CVSSP), University of Surrey. Haohe's research is mainly related to the topics on speech, music, and general audio, such as generative model, source separation, quality enhancement, and recognition. Haohe has served as a reviewer for various conferences and journals including INTERSPEECH, ICASSP,  TASLP, TNNLS, TCSVT, and ACM multimedia.
\end{IEEEbiography}
\vspace{-20pt}
\begin{IEEEbiography}[{\includegraphics[width=1in,height=1.25in,clip,keepaspectratio]{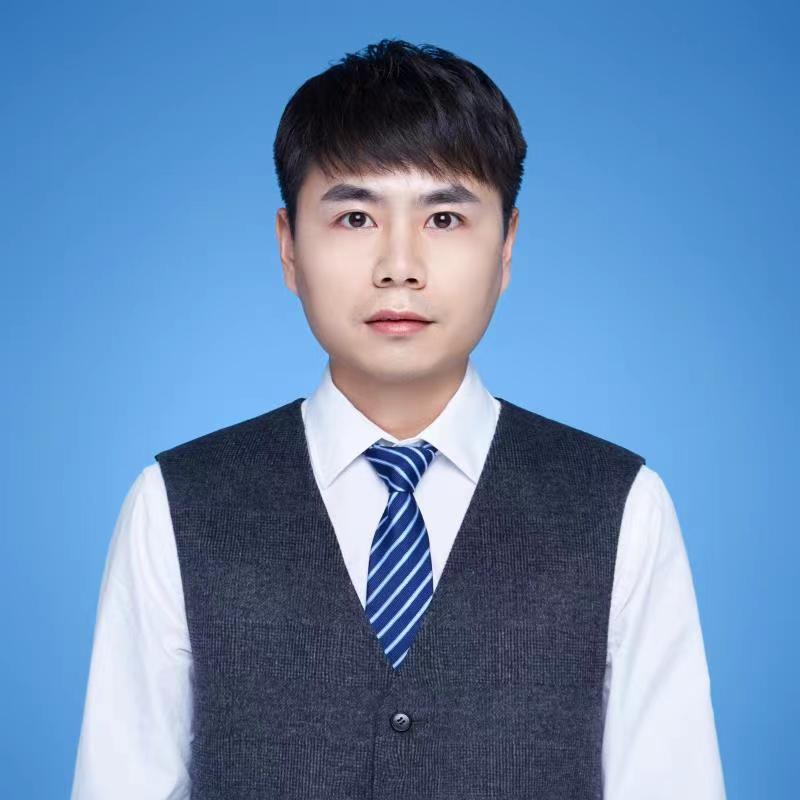}}]{Zhuangzhuang Du}
received the M.Sc. degree from the College of Agricultural Equipment Engineering, Henan University of Science and Technology, Luoyang, China, in 2021. He is currently pursuing the Ph.D. degree with the College of Information and Electrical Engineering, China Agricultural University, Beijing, China.
His research interests include wireless sensor network, sensor fault diagnosis research and intelligent algorithm development.
\end{IEEEbiography}
\vspace{-30pt}

\begin{IEEEbiography}[{\includegraphics[width=1in,height=1.25in,clip,keepaspectratio]{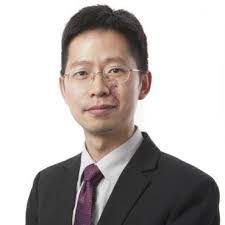}}]{Tao Chen}
received B.Sc., and M.E., degrees from Tsinghua University in China in 2002, and a Ph.D. in Chemical Engineering/System Engineering at Newcastle University. In 2011, he joined the University of Surrey as a lecturer and was promoted to the position of professor in August 2021. Currently, he serves as the Deputy Dean for International Collaboration at the School of Chemistry and Chemical Engineering at the University of Surrey. His current research focus is on the application of digital technologies in chemical engineering. 
\end{IEEEbiography}
\vspace{-30pt}

\begin{IEEEbiography}[{\includegraphics[width=1in,height=1.25in,clip,keepaspectratio]{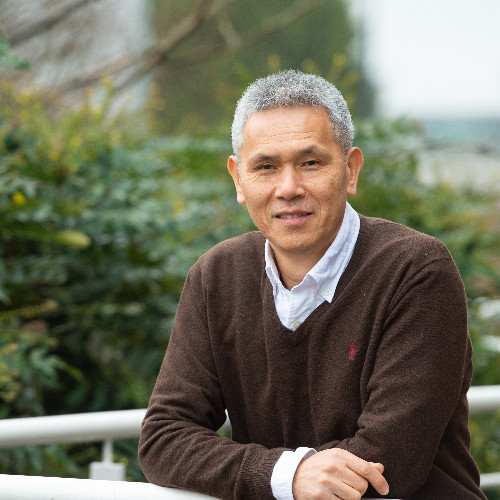}}]{Guoping Lian}
 is a Professor of Chemical and Process Engineering at the University of Surrey and Science Leader at Unilever R\&D Colworth in the UK. He received his BSc from Hunan Agriculture University, MSc from China Agriculture University, and PhD from Aston University in the UK. Although, primarily working in industrial R\&D, he has made significant contributions to scientific research. He spearheaded the application of multiphysics modelling and multiscale methods to investigate the complex multiphase materials and processes that are commonly encountered in the fine chemical, pharmaceutical, and food industries. Examples include the theoretical study of capillary force between nano/micro particles, DEM simulation of moist particulate systems, mathematical modelling of controlled release, and multi-physics modelling of crystallization, to name a few.
 \end{IEEEbiography}
\vspace{-30pt}
\begin{IEEEbiography}[{\includegraphics[width=1in,height=1.25in,clip,keepaspectratio]{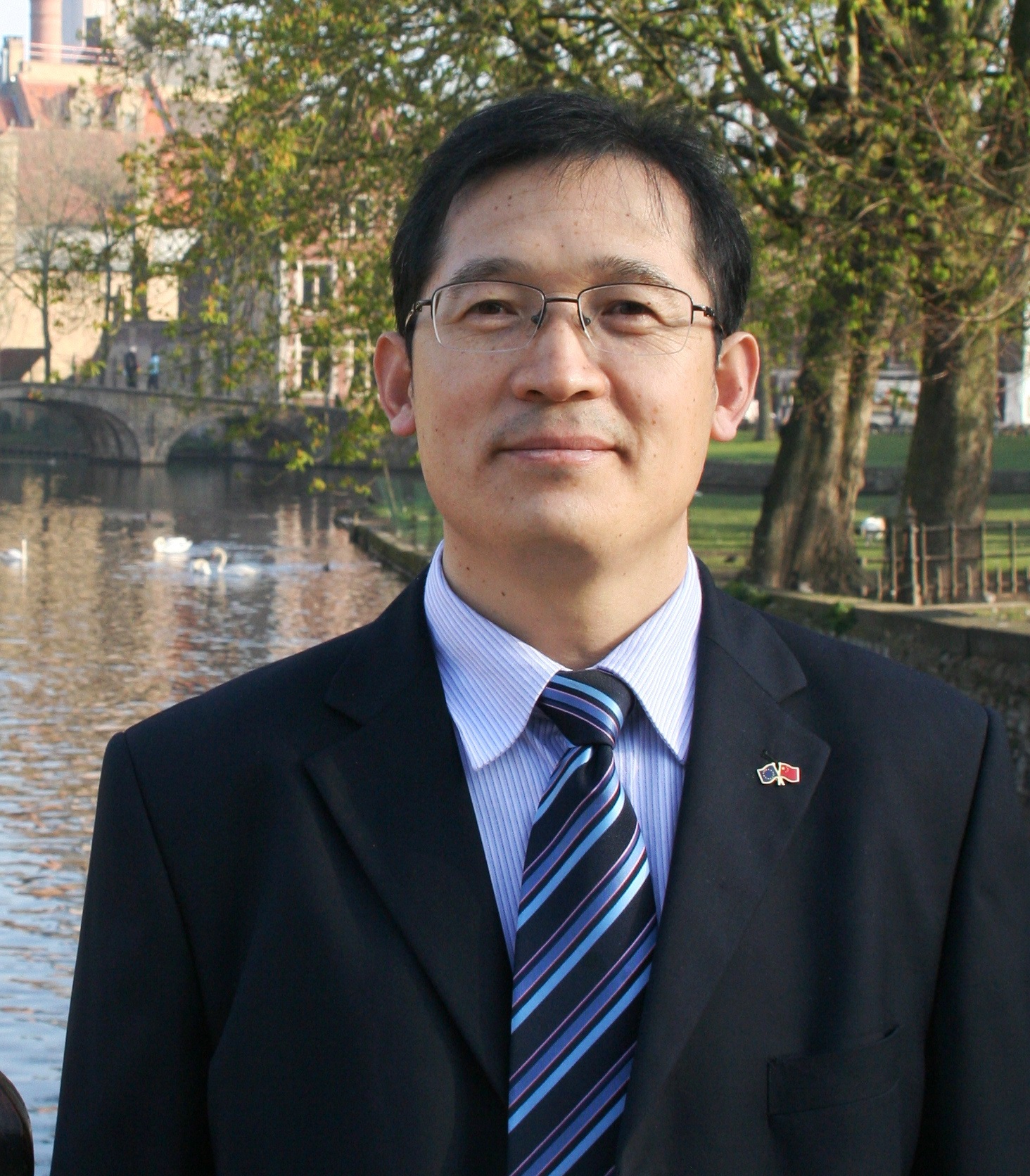}}]{Daoliang Li}
 is a Professor of College of Information and Electrical Engineering and the director of the National Innovation Center for Digital Fishery, China Agricultural University. He received the Ph.D. degree from the College of Engineering, China Agricultural University, his principal research interest is ICTs in aquaculture and agriculture, especially for information processing, smart sensors and smart control system in fish farming and aquaponics. He is the editor-in-chief of International Journal of Information processing in Agriculture  and the Chair of the Work Group for Advanced Information Processing in Agriculture, International Federation for Information Processing. He is also the deputy director for the of Expert Consultant Committee for National Agricultural and Rural informatization, Ministry of Agriculture and rural development. He was the chairman of of 1st to 13th International conference on computer and computing technologies in agriculture (www.iccta.cn). \end{IEEEbiography}
\vspace{-200pt}
\begin{IEEEbiography}[{\includegraphics[width=1in,height=1.25in,clip,keepaspectratio]{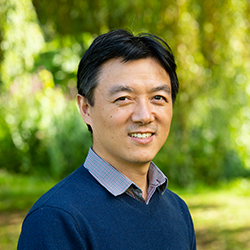}}]{Wenwu Wang}
was born in Anhui, China. He received the B.Sc., M.E., and the
Ph.D. degrees from Harbin Engineering University, China, in 1997, 2000, and 2002, respectively. He then worked at King’s College London, Cardiff University, Tao Group Ltd. (now Antix Labs Ltd.), and Creative Labs, before joining the University of Surrey,
U.K., in May 2007, where he is currently a Professor in signal processing and machine learning, and a Co-Director of the Machine Audition Lab within the Centre for Vision Speech and Signal Processing. He is also an AI Fellow at the Surrey Institute for People Centred Artificial Intelligence. His current research interests include signal processing, machine learning and perception, artificial intelligence, machine audition (listening), and statistical anomaly detection. He has (co)-authored over 300 papers in these areas. He has been recognized as a (co-)author or (co)-recipient of more than 15 awards, including the 2022 IEEE Signal Processing Society Young Author Best Paper Award, ICAUS 2021 Best Paper Award, DCASE 2020 and 2023 Judge’s Award, DCASE 2019 and 2020 Reproducible System Award, and LVA/ICA 2018 Best Student Paper Award. He is an Associate Editor (2020-2025) for IEEE/ACM Transactions on Audio Speech and Language Processing. He was a Senior Area Editor (2019-2023) and Associate Editor (2014-2018) for IEEE Transactions on Signal Processing. He is the elected Chair (2023-2024) of IEEE Signal Processing Society (SPS) Machine Learning for Signal Processing Technical Committee, a Board Member (2023-2024) of IEEE SPS Technical Directions Board, the Chair (2025-2027) and Vice Chair (2022-2024) of the EURASIP Technical Area Committee on Acoustic Speech and Music Signal Processing, an elected Member (2021-2026) of the IEEE SPS Signal Processing Theory and Methods Technical Committee. He was a Satellite Workshop Co-Chair for INTERSPEECH 2022, a Publication Co-Chair for IEEE ICASSP 2019, Local Arrangement Co-Chair of IEEE MLSP 2013, and Publicity Co-Chair of IEEE SSP 2009. He is a Satellite Workshop Co-Chair for IEEE ICASSP 2024, Special Session Co-Chair of IEEE MLSP 2024, and Technical Program Co-Chair of IEEE MLSP 2025.

\end{IEEEbiography}

\end{document}